# Dynamical thickening transition in plate coating with concentrated surfactant solutions

Jérôme Delacotte[1], Lorraine Montel[1], Frédéric Restagno[1], Benoît Scheid[2], Benjamin Dollet[3], Howard A. Stone[4], Dominique Langevin[1], and Emmanuelle Rio[1]

[1] Laboratoire de Physique des Solides UMR8502, Université Paris-Sud - 91405 Orsay, France, EU
[2] TIPs - Fluid Physics unit, Université Libre de Bruxelles - C.P. 165/67, 1050 Brussels, Belgium, EU
[3] Institut de Physique de Rennes UMR6251, Université de Rennes - 35042 Rennes, France, EU
[4] Department of Mechanical and Aerospace Engineering, Princeton University, Princeton, NJ 08544, USA

**Abstract**

We present a large range of experimental data concerning the influence of surfactants on the well-known Landau-Levich-Derjaguin experiment where a liquid film is generated by pulling a solid plate out of a bath. The thickness $h$ of the film was measured as a function of the pulling velocity $V$ for different kind of surfactant and at various concentrations. Measuring the thickening factor $\alpha = h/h_{\mathrm{LLD}}$, where $h_{\mathrm{LLD}}$ is obtained for a pure liquid, in a wide range of capillary ($Ca = \eta V/\gamma$), two regimes of constant thickening can be identified: at small capillary number, $\alpha$ is large due to a confinement and surface elasticity (or Marangoni) effects and at large $Ca$, $\alpha$ is slightly higher than unity, due to surface viscous effects. At intermediate $Ca$, $\alpha$ decreases as Ca increases along a “dynamic transition”. In the case of non-ionic surfactants, the dynamic transition occurs at a fixed $Ca$, independently of the surfactant concentration, while for ionic surfactants, the dynamic transition depends on the concentration due to the existence of an electrostatic barrier. The control of physico-chemical parameters allowed us to elucidate the nature of the dynamic transition and to relate it to surface rheology.

# 1. Introduction

When a solid object is pulled out of a liquid reservoir, a thin layer of liquid is entrained by viscous drag. Since coating flows are ubiquitous in industrial processing, understanding the variables that control the film thickness is of major importance. In industrial processes, the coatings can be made of pure liquid such as oils, but can also be paints, emulsions, polymers solutions, i.e. complex fluids. The coating materials protect, functionalize, and lubricate surfaces. In most cases, it is desirable to obtain a well controlled thickness of the applied layer and a high coating speed to maintain a high throughput. Therefore it is of interest to determine the dependence of thickness of these thin films on physico-chemical parameters. In this paper we report experimental results of solid plates coating by various types of surfactant solutions in a wide range of concentrations above the cmc. Our results are compared to available theoretical models.

The classic film-coating theory by Landau-Levich and Derjaguin (LLD) uses the lubrication and low capillary number approximations and then solves the governing equations by matching the thin film region (of constant thickness $h$) far away from the bath with the static meniscus (near the horizontal bath) through an intermediate transition region called the "dynamical meniscus" of length $\ell$ (Levich 1962), as illustrated in Figure 1. The calculation is based on an asymptotic matching approach, and a numerical calculation is used to obtain the film thickness:

$$h_{LLD} = 0.9458\, \ell_c\, Ca^{2/3}, \qquad (1)$$

where $Ca = \eta V/\gamma$ is the capillary number, $V$ the plate velocity, and $\eta$ and $\gamma$ respectively the viscosity, and the surface tension of the liquid into air; $\ell_c = \sqrt{\gamma/\rho g}$ is the capillary length, with $\rho$ the density of the liquid. Physically, equation (1) means that increasing the velocity of the plate or the viscosity of the liquid leads to an increase of the drag force, and then to a thicker film. On the other hand, increasing the surface tension of the liquid decreases the film thickness. This calculation imposes a no-slip boundary condition at the solid-liquid interface and zero tangential stress at the liquid-air interface. The LLD calculation is valid for $Ca^{1/3} \ll 1$ since when the capillary number is close to one, gravitational effects cannot be neglected (Derjaguin 1943) (note that this condition becomes $Ca^{1/3}Bo \ll 1$ in cylindrical geometries). We will work in the small $Ca$ regime in the present paper. For a liquid of viscosity $\eta$ = $10^{-3}$ Pa.s, a surface tension $\gamma$ between 30 and 40 mN/m, and a density $\rho$ = $10^3$ kg/m$^3$, the film thickness is predicted to vary from 17 nm to 20 μm, if $V$ varies from 1 μm/s to 40 mm/s ( $2.5\times10^{-8} < Ca < 10^{-3}$ ).

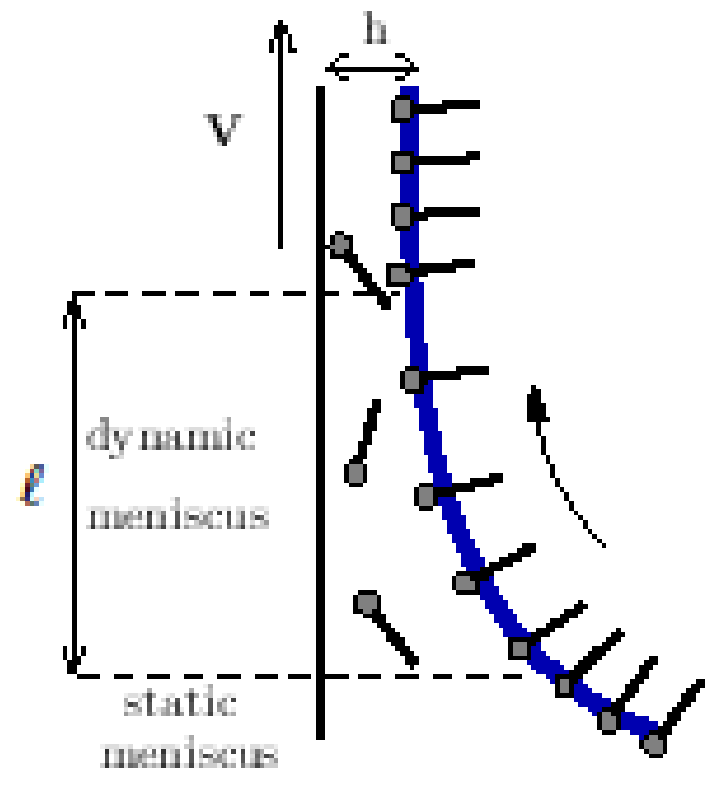

**Figure 1: Velocity controlled withdrawal of a solid plate out of a liquid bath. The air-liquid interface is stretched in the so-called dynamic meniscus, which has an extension $\ell$ and connects the static meniscus with the flat film region. In the dynamic meniscus, the viscous drag is balanced by surface tension forces. In the case of a surfactant solution, Marangoni effects and interfacial viscous effects can also be important.**

There are only a few experimental validations of the LLD law for simple liquids. In a planar configuration. Krechetnikov and Homsy (2005, 2006) measured the thickness of the liquid films by measuring the weight of the entrained liquid. Using glycerol-water solutions over a wide range of $Ca, 10^{-3} < Ca < 10^{-1}$, they reported agreement with equation (1), with small corrections for a finite bath size and an overall accuracy of 10 %. In particular, the power law of ⅔ was verified with an accuracy of 5%. Note that in the case of pure liquids with a substrate of controlled roughness, they observed a significant thickening of the films relative to those on smooth substrates and a different power of capillary number than predicted in equation (1). More recently, Snoeijer et al. (2008) reported a precise validation of the LLD law for silicon oil and a wetting surface. The thickness of the film was measured using an interferometric technique, and the precision was as good as 0.2 %. We note that the authors also showed that in the case of partial wetting, another solution different than the LLD law exists, which has a larger film thickness and scales as $\ell_c\, Ca^{1/2}$.

Most coating flow experiments involving withdrawal of a substrate have been performed on curved surfaces such as the coating of fibers () or capillary tubes (Chen 1986, Schwartz et al. 1986). The thickness of the withdrawn film is then given by the Bretherton law, which is a variant of the LLD law (Bretherton 1961). Indeed, since the radius of curvature in the static meniscus is directly given by the tube radius $r$ rather than by $\ell_c/\sqrt{2}$ (if $r \ll \ell_c$), the Bretherton law is $h = 1.34\, r\, Ca^{2/3}$. The film thickness in fiber coating obeys the same law, taking $r$ for the fiber radius.

.

Several studies have indicated deviations from the LLD law when complex fluids are used. We will only recall here the experiments made with solutions containing surface-active substances. In the plate drag-out problem, Groenveld used water-glycerol solutions containing traces of hexane or oil and measured films thicker than predicted by the LLD law (Groenveld 1970). Recently, Krechetnikov and Homsy (2005) reported experiments using sodium dodecyl sulfate (SDS) solutions whose critical micellar concentration (cmc, concentration above which micelles form) is 8.3 mM. They observed that the film tends to thicken when a surfactant is added (the ratio $c$/cmc was varied between 0.2 and 1). They defined, as other authors in other geometries, a thickening factor $\alpha$ which is the ratio of the measured film thickness $h$ to the film thickness predicted by the LLD relation:

$$\alpha = \frac{h}{h_{LLD}} \ . \tag{2}$$

The theoretical thickness $h_{LLD}$ was calculated by using in equation (1) the surface tension of the surfactant solution $\gamma$ measured at equilibrium at the specified bulk concentration. For a Newtonian fluid, $\alpha$=1; Krechetnikov and Homsy (2005) found $\alpha$ = 1.55, independently of $Ca$ in the range $10^{-4}$-$10^{-3}$ and of the surfactant concentration. Note finally that Krechetnikov and Homsy (2006) numerically predicted thinning of the withdrawn film, in qualitative disagreement with the experiments. However, systematic values of the sorption rates would be required to check whether such a thinning is general. Moreover, Campana et al. (2010) proposed a numerical simulation that predicts a thickening due to surfactant and could rationalize the experimental results of Krechetnikov and Homsy (2005). It remains therefore unclear what is the role of the sorption barriers in the general case.

The largest amount of experimental data concerns fiber coating. To our knowledge all the experiments have shown a thickening due to surfactants. For this geometry, the thickening factor $\alpha$ has been shown to depend on the chemical nature of the surfactant, on the surfactant concentration, on the radius of the fiber, and on the capillary number. The most extensive study has been done with SDS solutions (Quere 1998, 1999): the thickening factor ($\alpha$) was measured as a function of concentration, it was observed to increase before the cmc and reach a maximal value of 2.2 around the cmc, and eventually to decrease to a constant value of 1.6 between 1 and 10 cmc. In the same paper, a single set of experiments is reported with a different surfactant, dodecyl trimethyl ammonium bromide (DTAB). A dynamic thickening transition is then observed by increasing the capillary number (increasing the withdrawal velocity), after which the thickening factor $\alpha$ decreased toward unity. In this article, we choose to investigate this dynamical transition of thickening. We then did systematic experiments with concentrated surfactants both ionic and non ionic, which is of large importance, as it will be stated in the following.

Note that many other experiments exhibiting data far away from the LLD (or Bretherton) power-law are available in the literature, most of them dealing with the bubble-in-a-tube or fiber geometry. Many different mechanisms can be invoked for such a deviation: gravity effects appear at high capillary numbers (White 1965), the circular shape intrinsic to the bubble experiment can also lead to deviations at small capillary numbers (Schwartz 1986). Last but not least, such a transition has also been observed in the bubble-in-a-tube geometry, using pure liquids (Bretherton 1961, Chen 1986), where it is suggested that a very small amount of surfactants, leading to strong gradients, is at the origin of this transition. None of these mechanisms can be invoked to explain our experiments since we are working with concentrated surfactants, at small capillary numbers and in a flat geometry.

The thickening is usually ascribed to the Marangoni effect, i.e. a surface concentration gradient leading to an additional stress at the surface. This Marangoni effect can be made quantitative by introducing the surface elasticity. The surface compression elastic modulus, $E$, is defined as:

$$E = -A\frac{d\Pi}{dA} \ ,$$

where $A$ is the surface area, $\Pi = \gamma_w - \gamma$ the surface pressure, with $\gamma_w$ the surface tension of pure water. $E$ is interpreted as the 2D analog of a 3D compression bulk modulus $-\mathcal{V}dp/d\mathcal{V}$, where $p$ is the pressure and $\mathcal{V}$ the volume. For insoluble surfactants, $E = \Gamma \frac{d\Pi}{d\Gamma}$, with $\Gamma$ the surface concentration of surfactant. The surface compression process can be accompanied by friction in the surface layer, in which case, a two-dimensional compression surface viscosity $\eta_E$ is introduced. Note that shear can also be applied, in which case, a shear modulus $S$ and a shear viscosity $\eta_S$ need to be introduced.

When soluble surfactants are used, the problem is more complicated since exchanges between surface and bulk are possible. When the time scale of the compression is comparable to the exchange time between the bulk and surface, the resistance to compression is lowered and the apparent elastic modulus $E$ is smaller than the value only due to the monolayer at the interface. There is a significant dissipation associated with the surface-bulk exchange, and the apparent surface viscosity $\eta_S$ is much larger than the value intrinsic to the monolayer (Stevenson 2005). Elasticity as well as viscosity then depends not only on the surfactant concentration but also on the perturbation time scale of the interface. The values of E and $\eta_S$ tend to decrease significantly at high concentration and/or at low frequency since surfactants have time to exchange between surface and bulk. This variation has been modeled by Levich (1962) and by Lucassen and van den Tempel (1972). Note also that soluble surfactant monolayers usually flow under shear stresses, for which we expect that the shear elastic modulus S is zero and the shear viscosity $\eta_S$ is small. Stebe et al. studied the remobilization of surfaces in a three-phase slug flow (Stebe 1991).They showed that at high surfactant concentration or at a high transport rate of surfactants, a uniform concentration of surfactant at the interfaces leads to a response similar to a clean interface (with a lower surface tension). There has been an important theoretical effort to model the effect of an interfacial stress at the interface and its consequence in a coating experiment. Most of this effort has been concentrated on the Bretherton geometry. Park (Park 1991) and Chang and Ratulowski (1990) studied the deviation from Bretherton's results in the case of small amounts of surfactants. Stebe and Barthès-Biesel (1995) studied the case of elevated surfactant concentrations. All of these theoretical works conclude that when the surface elasticity and/or viscosity increase there exists an upper bound on the film thickness, which is $4^{2/3}$ times the value obtained by Bretherton with pure liquids.

In the case of coating processes, another feature to be considered is the film thickness: if the film is too thin, it cannot act as a reservoir of surfactant for the surface (Lucassen 1981). Quéré and de Ryck introduced the dimensionless number $\sigma = \Gamma/(ch)$, which compares the amount of surfactant at the surface with the amount of surfactant in the film, where $\Gamma$ is the surface concentration and $c$ is the bulk concentration. If $\sigma >> 1$, there is not enough surfactant in the film to replenish the surface during film stretching; in this case, the elastic modulus $E$ is much larger than that of the surface of a solution with the same surfactant concentration, and is rather close to the intrinsic elastic modulus $E_0$ (Lucassen 1981). Note that Quéré and de Ryck attributed the observed dynamical transition of thickening to the transition between regimes with $\sigma>>1$ and $\sigma<<1$. However, in their experiments the transition occurred around $\sigma=10^{-2}$ and not unity. We will discuss this apparent contradiction below.

It is always difficult to separate the role of surface viscosities and elasticities. The film response is never purely elastic as assumed in the models based on Marangoni effects. It is never purely viscous either. In a previous paper (Scheid 2010), we chose particular experimental conditions to be only sensitive to the intrinsic surface viscosity effects. We showed that at high concentrations of surfactant and at capillary numbers above the dynamical transition of thickening, a thickening effect of about 6% was observed with DeTAB, which could thus be rationalized by the sole effect of the intrinsic surface viscosity.

Despite the present knowledge on coating problems, many important questions regarding coating with surfactant solutions still remain to be understood. For example, little is known about the film thickness variation with surfactant concentration well above the critical micellar concentration and about the role of surfactant solubility and the surface rheology on the film thickness. In this article, we focused on this situation (role of surface rheology at large surfactant concentration) and describe a set of systematic experiments with several surfactant solutions. We study the film thickening properties of a non-ionic surfactant, $C_{12}E_6$, and two cationic surfactants, DeTAB and DTAB. A comparison between the experimental results and the existing theories is made.

# 2. Experiments

## *2.1 Apparatus and methods*

We built an experimental set-up that allows controlling the film formation and measuring its thickness. A translation stage (Newport UTS 150CC) coupled with a controller (Newport SMC100CC) was used to drive a bath of solution down with a controlled velocity (1 μm/s-40 mm/s ± 1μm/s). The film thickness was measured using an interferometric technique. A white light beam is reflected by both interfaces of the film and analyzed using a spectrometer. Both the light source (LS-100) and the spectrometer (USB 400) are Ocean Optics devices. The range of wave lengths span from 450 to 900 nm. The apparatus is shown in Figure 2. Silicon wafers (Siltronix 111) were used as solid plates for the withdrawal experiments. They exhibit low roughness at the atomic scale and were cleaned, just before each experiment, using both piranha solution and UV-ozone cleaner to ensure good wettability. To avoid any edge effects, the measurements are made in the middle of the wafer which has a size around 5 cm in each direction. The thickness is measured above the dynamic meniscus, at a distance around two times the capillary length from the horizontal surface in order to be in the flat zone of the entrained film.

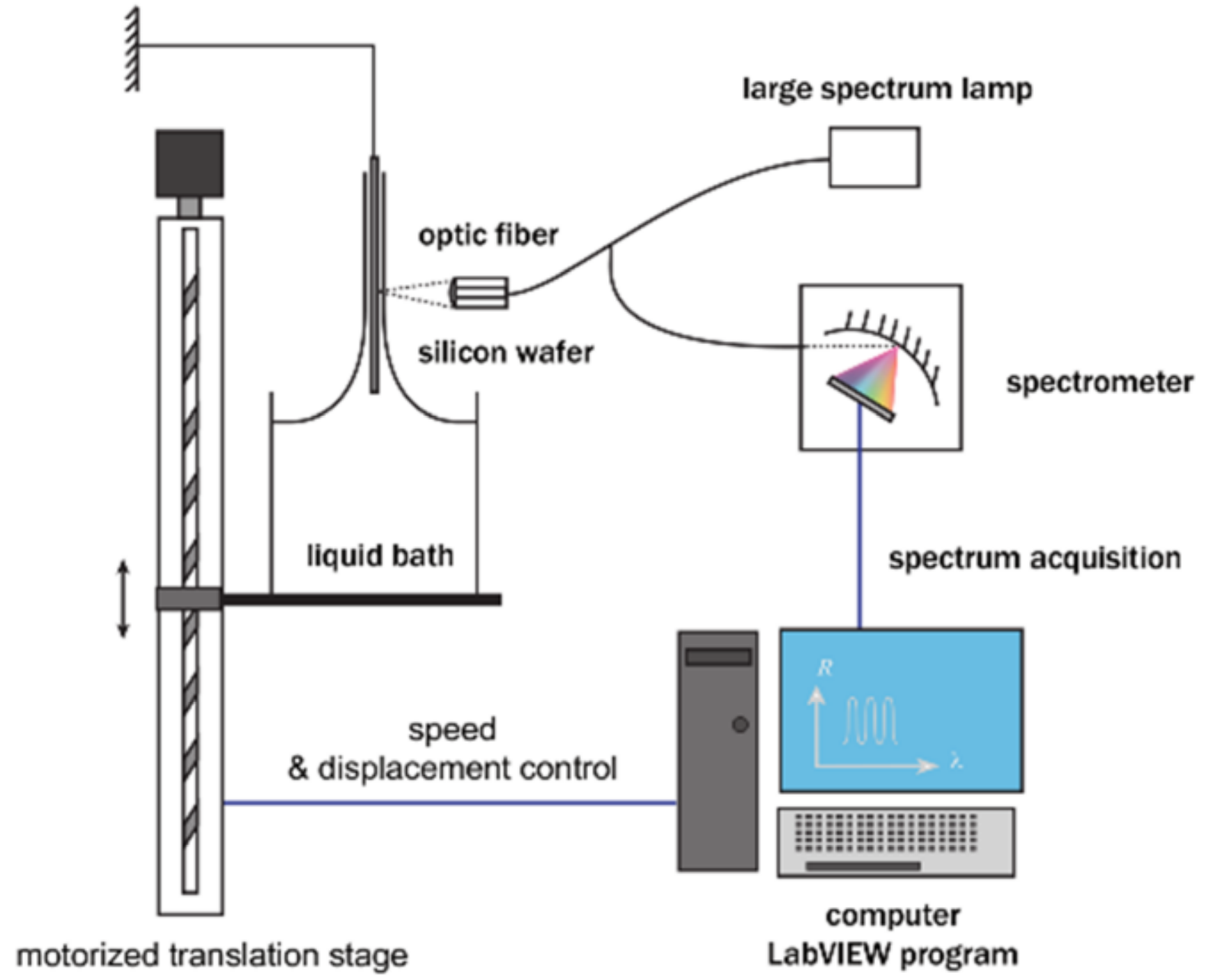

**Figure 2: Experimental set-up for film coating. A translation stage moves the bath of solution with a controlled velocity. The plate is coated by a liquid film whose thickness is measured using a spectrometer. The reflectivity is recorded as a function of the wavelength of light.**

The raw data, i.e. the reflectivity *R* as a function of the wavelength, was monitored with the Spectrasuite interface software from Ocean Optics. The film thickness was determined, with 5% accuracy, by adjustment of the data using the following expression (Born and Wolf 1985):

$$R = b\frac{\left(\frac{n^2-1}{2n}\right)^2 \sin^2\left(2\pi\frac{nh}{\lambda}\right)}{1+\left(\frac{n^2-1}{2n}\right)^2 \sin^2\left(2\pi\frac{nh}{\lambda}\right)} + d\,, \tag{3}$$

where $h$, $\lambda$ and $n$ are, respectively, the film thickness, the wavelength and the refractive index of the solution. The latter was measured with a refractometer (OPL) after each experiment. No correction is required for the presence of the surfactant monolayers (whose thickness is of the order of 1nm) since *h* is of the order of microns. Fitted parameters are *b*, *d* and *h*. Note that *b* and *d* could have been expressed as functions of the refractive indices of the silicon wafer and the solution. We used them as fitting parameters to make the procedure simpler. In order to make the adjustment less sensitive to the initial value of $h$, some researchers use a simple cosine function (Snoeijer 2008). Figure 3 shows an example of the thickness determination for a surfactant solution.

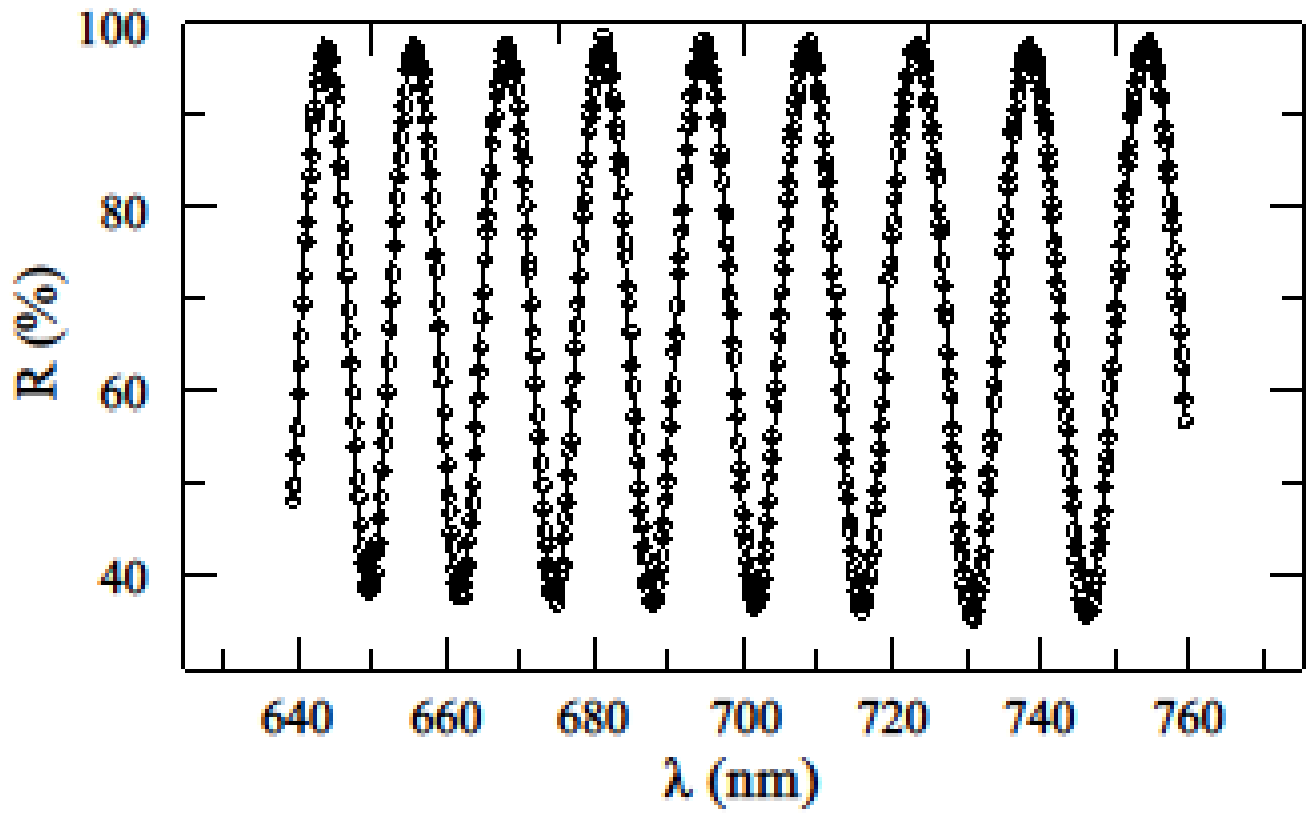

**Figure 3 : Reflectivity spectrum recorded from the spectrometer with the Spectrasuite software (dots) and fit with equation (3) (line) for a film made of a 990 mM DeTAB solution. The film thickness obtained in this example is *h*=12.7 μm, and other parameters are *b*=640.1, *d*=38.1, with *n*=1.374.**

The validation of the experimental set-up was made with a pure liquid, whose properties are easy to control. We chose a silicon oil (Rhodorsil 47V20) instead of water, since the surface of water is difficult to keep uncontaminated after several withdrawals of a plate, even with careful handling and filtration. Figure 4 shows the film thickness scaled by the capillary length $\ell_c$ as a function of the capillary number $Ca = \eta V / \gamma$ ; we use logarithmic scales, throughout this paper.

In this figure, the dashed line corresponds to the mean value for the thickening factor, $\alpha$ = 0.99, which is 1% below the theoretical value, lying in turn within the standard deviation of 2%. The error on the thickness measurement was then evaluated at a maximum value of 5% (including reproducibility), which is beyond the size of the experimental points in all of the figures.

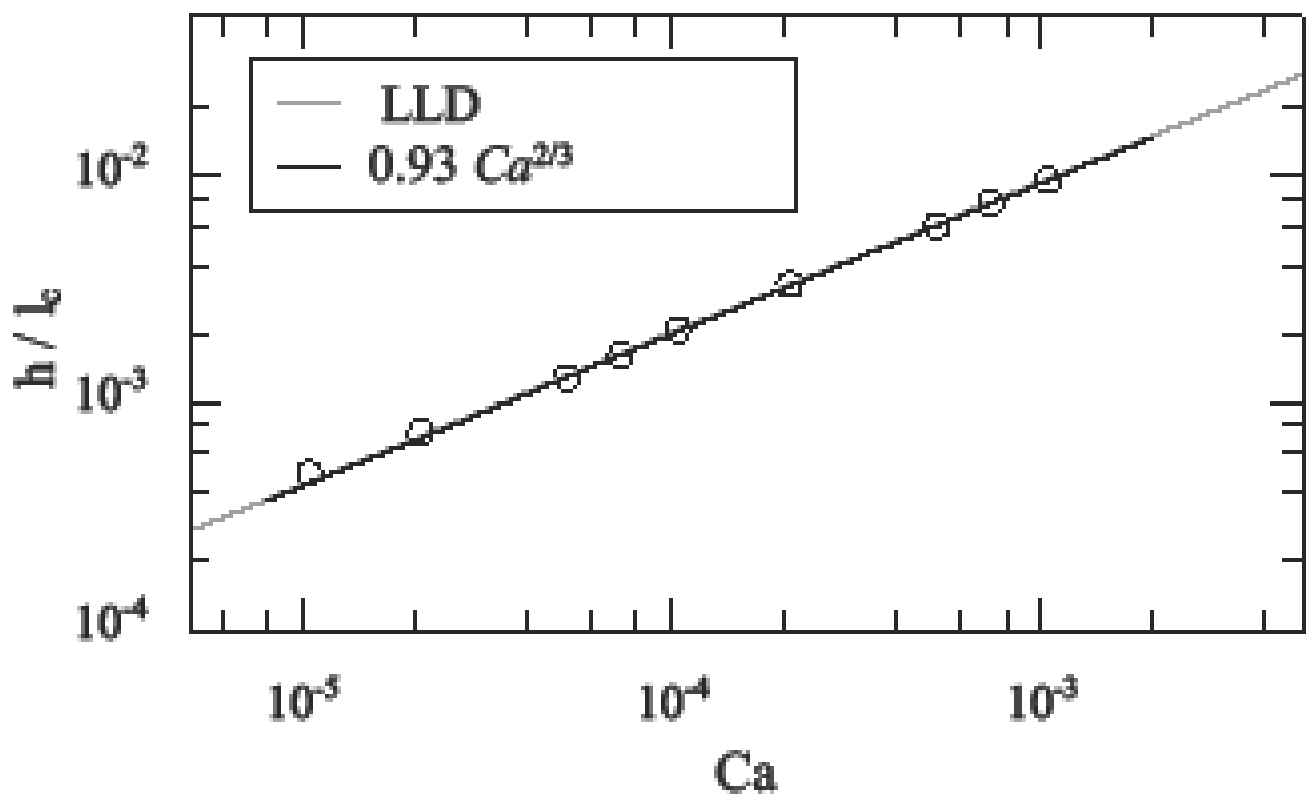

**Figure 4 : Validation of the experimental set-up by using a silicon oil 47V20 ( $\eta$ = 20 mPa.s, $\gamma$ = 21 mN/m). The film thickness $h$ is well predicted by the LLD model, equation (1).**

## 2.2 Materials

We used three different surfactants in order to vary the distribution of surfactant molecules between the bulk and the surface of the film: dodecyl trimethyl ammonium bromide ( $D$TAB ), decyl trimethyl ammonium bromide ( $De$TAB ) and hexaethyleneglycol-monododecylether ( $C_{12}E_6$ ). Their critical micellar concentrations (cmc) are given in table 1. $D$TAB was purchased from Sigma-Aldrich and was recrystallized three times before use in order to decrease the amount of impurities. $De$TAB (purity 99 %) and $C_{12}E_6$ (purity 98 %) were purchased from Sigma-Aldrich and used as delivered. Note that DTAB is very stable against hydrolysis, and, to avoid chemical decomposition and solution aging, we always used freshly prepared solutions of $C_{12}E_6$, which hydrolyses only very slowly (on the time scale of a week). Water used in the experiments was ultra-purified water from a Millipore-Q instrument (resistivity = 18 MΩ cm).

| surfactant | *cmc* (mM) | $\gamma_{cmc}$ (mN/m) | $\ell_c$ (mm) |
|---|---|---|---|
| $C_{12}E_6$ | 0.07 | 32.3 | 1.82 |
| $D$TAB | 15 | 38.0 | 1.97 |
| $De$TAB | 66 | 39.7 | 2.01 |

**Table 1: Critical micellar concentration (cmc) of the three surfactants used in this study, together with the surface tension measured at the cmc (with an experimental error of ±O.5 mN/m), and the capillary length calculated with the density of the solution at the cmc.**

Surfactant $C_{12}E_6$ has the lowest monomer solubility (corresponding to the lowest cmc) since it is nonionic (Durbut 1999). The viscosities of all surfactant solutions were measured with a low shear rheometer (Low Shear 30 Contraves) at a temperature of 25°C at which the experiments have been conducted. Viscosities of $C_{12}E_6$ solutions were comparable to that of water for the all range of surfactant concentration. Measured viscosities for ionic surfactants (DTAB and DeTAB) reached somewhat larger values (up to 3.35 mPa.s) when the concentration was increased, which are still low enough to exclude the presence of wormlike micelles or liquid crystalline phases in the bulk (Israelachvili 1992). The surface tensions were measured for all solutions using a Wilhelmy plate apparatus, with an accuracy of 0.5 mN/m.

As mentioned in the introduction, we are working at very small capillary numbers for which gravity is negligible, i.e. $Ca^{1/3} \ll 1$. Moreover, with a maximum velocity of 40 mm/s, the Weber number that compares inertial to capillary effects is We=$\rho V^2 l_c/\gamma \approx 10^{-1} \ll 1$. Hence inertial effects are also negligible here."

# 3. Results and discussion

In this section, we present the results obtained by varying the type of surfactant and the concentration. The film thickness was measured as a function of the capillary number and concentration.

## *3.1 Non-ionic surfactant*

The concentrations of $C_{12}E_6$ solutions spanned from 0.07 to 3.5 mM (i.e., in the range 1 – 50 cmc). Figure 5 shows the variation of film thickness ($h$) rescaled by the capillary length ($\ell_c$) as a function of the withdrawal velocity ($V$) expressed in terms of the dimensionless capillary number ($Ca$). At small concentration (0.21 mM, or 3cmc, as shown in Figure 5(a)), a constant thickening factor $\alpha$ is observed in the range of investigated $Ca$. Indeed, the film thickness varies as the 2/3 power of the velocity. At these concentrations (Figure 5(a)), $\alpha$ is close to $\alpha_{max} = 4^{2/3}$. At higher concentrations (0.28 mM or 4 cmc and above, see Figure 5(b) - Figure 5(f)), we still observe a constant $\alpha$ for low thicknesses (i.e. low capillary numbers). However, when the film thickness increases, a transition occurs: $h \propto Ca^{2/3}$ is no longer observed and the thickening factor decreases. In this transition regime, the thickness variation as a function of the velocity agrees well with another power law $h \propto Ca^{½}$, as noted by Ou Ramdane and Quéré (1998). Whatever the concentration, the transition occurs around the same capillary number $Ca \approx 3\times10^{-5}$, which is associated with a film thickness $h \approx 3.8$ µm.

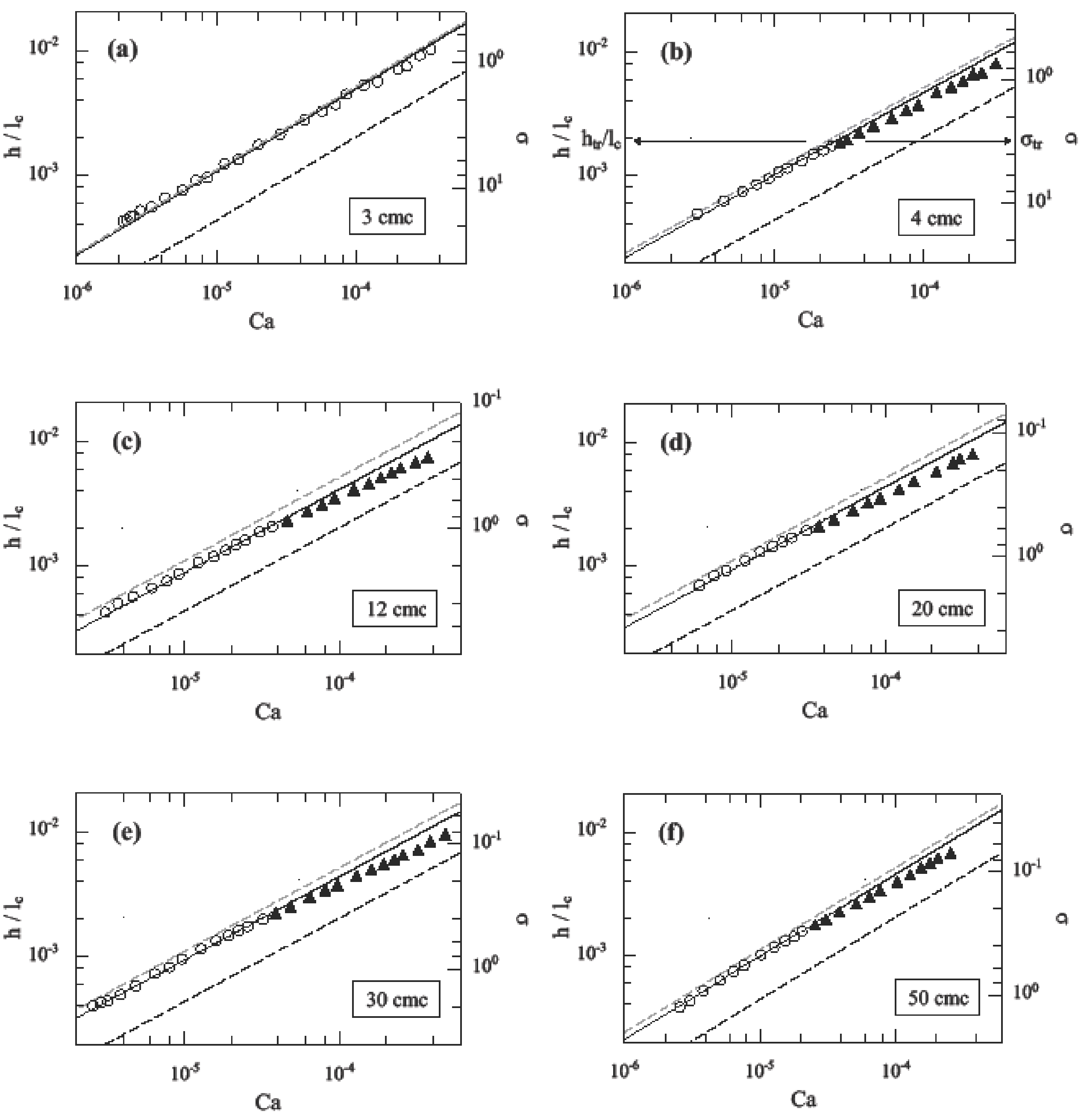

**Figure 5: Film thickness rescaled by the capillary length plotted as a function of the capillary number for various concentrations. The results correspond to various $C_{12}E_6$ concentrations in the solution (with a cmc of 0.07 mM). The right vertical axis shows the value of $\sigma$ (increasing from top to bottom) corresponding to each thickness. The dashed lines, bold and normal, show respectively the LLD thickness ($\alpha$=1) and the maximum possible thickness ($\alpha=4^{2/3}$) corresponding to an immobile interface. Solid line is a fit over the constant thickening region from which the value of α is obtained. Open and filled symbols represent data in the constant thickening region and in the thickening transition region, respectively. The subscript "tr" denotes parameters taken at theintersection between these two regions as represented in (b).**

The values of constant thickening α found at low capillary number are shown in figure 6 for each concentration: $\alpha$ first increases with concentration and then saturates above 0.21 mM (3 cmc) to a value α≈2.1.

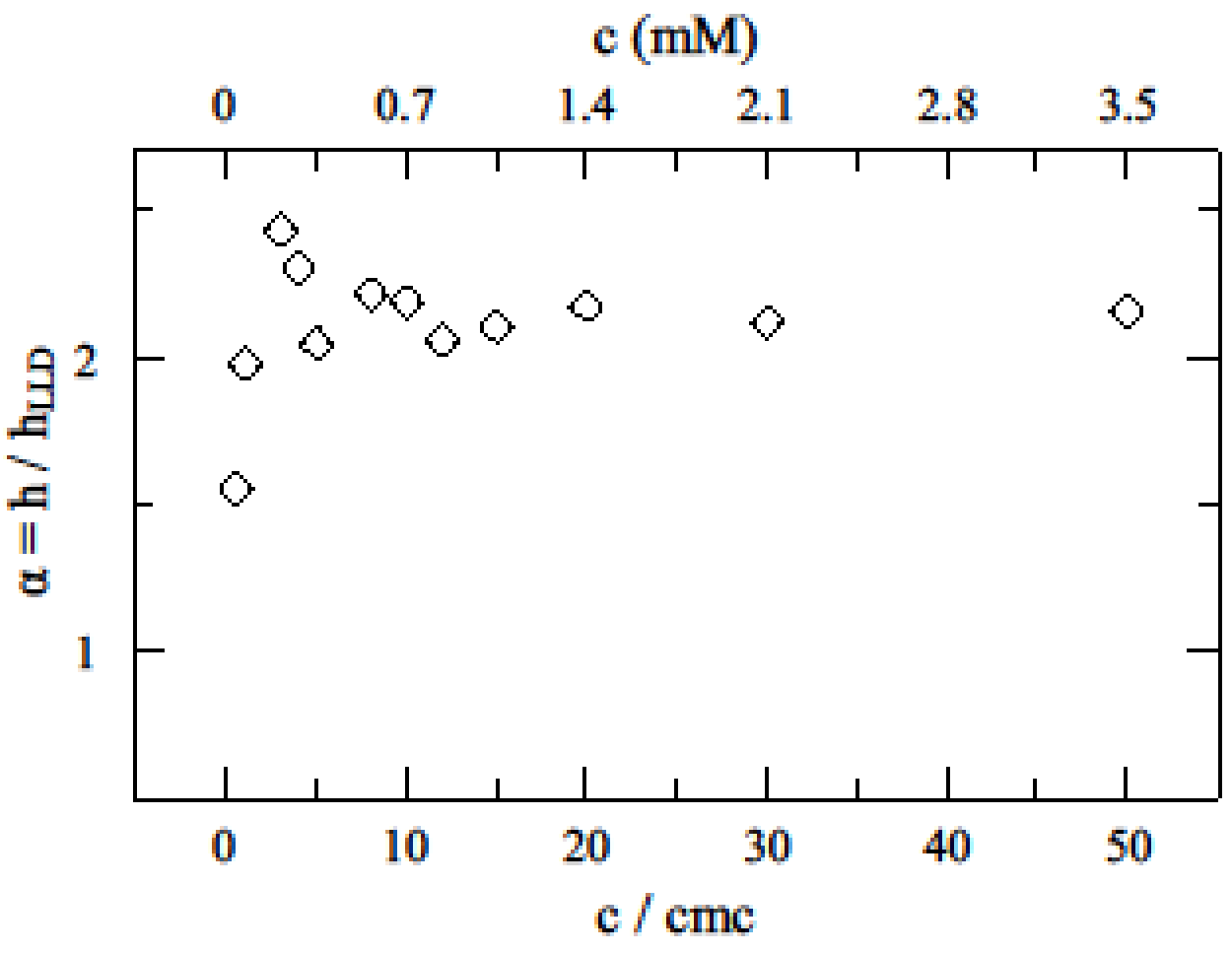

**Figure 6 : Thickening factor $\alpha$ versus $C_{12}E_6$ concentration rescaled by the cmc (0.07 mM).**

In order to reach higher concentrations without having elongated micelles or liquid crystalline phases (and larger viscosities that could vary with the velocity *V*, thus complicating the interpretation), we studied short chain cationic surfactants, for which we next report the results.

## *3.2 Ionic surfactants*

### 3.2.1 DTAB

The DTAB concentration was varied in a range spanning from 10 to 375 mM (i.e. 2/3 to 25 cmc). The experimental trends are qualitatively similar to those with $C_{12}E_6$. As shown in Figure 7(a) - Figure 7(d), for concentrations up to 150 mM (10 cmc), $\alpha$ remains constant when *Ca* is varied. At higher concentrations a transition in film thickening is observed: the thickening factor decreases toward values close to the LLD prediction (see Figure 7(e) - (f)). The transition occurs above $Ca \approx 10^{-4}$ and $h \approx 6$ µm. These values are similar to those obtained with $C_{12}E_6$ ($3\times10^{-5}$ and 4 µm, respectively).

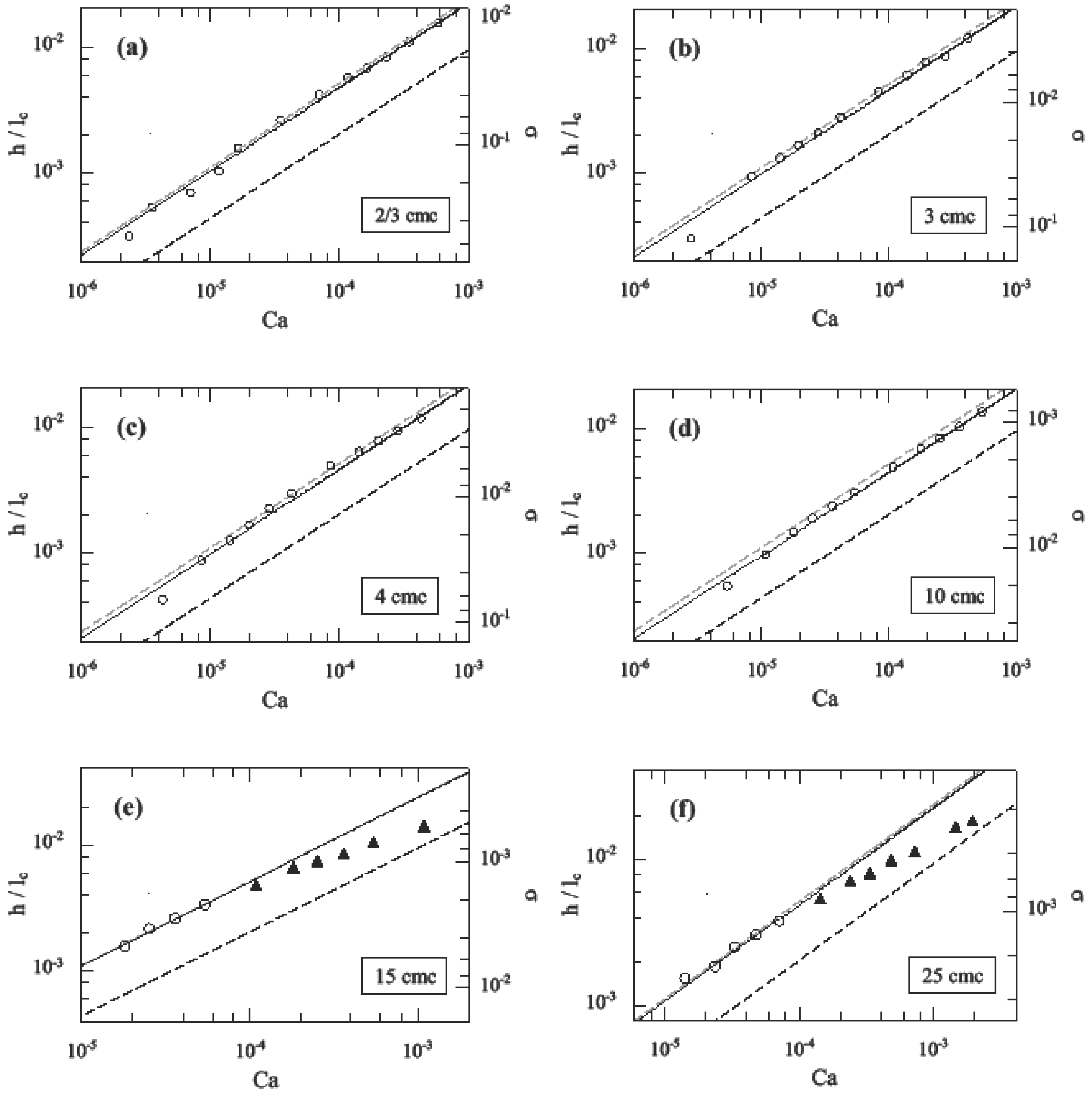

**Figure 7: Results obtained with DTAB for various concentrations (with a cmc of 15 mM). Same axes and notations as for Figure 5.**

In Figure 8, we plot the thickening factor variation with surfactant concentration at low *Ca*. As for $C_{12}E_6$, the thickening factor, for higher concentrations is the constant value obtained before the thickening transition (i.e. for small capillary numbers). Here $\alpha$ remains high and nearly constant when the concentration is varied.

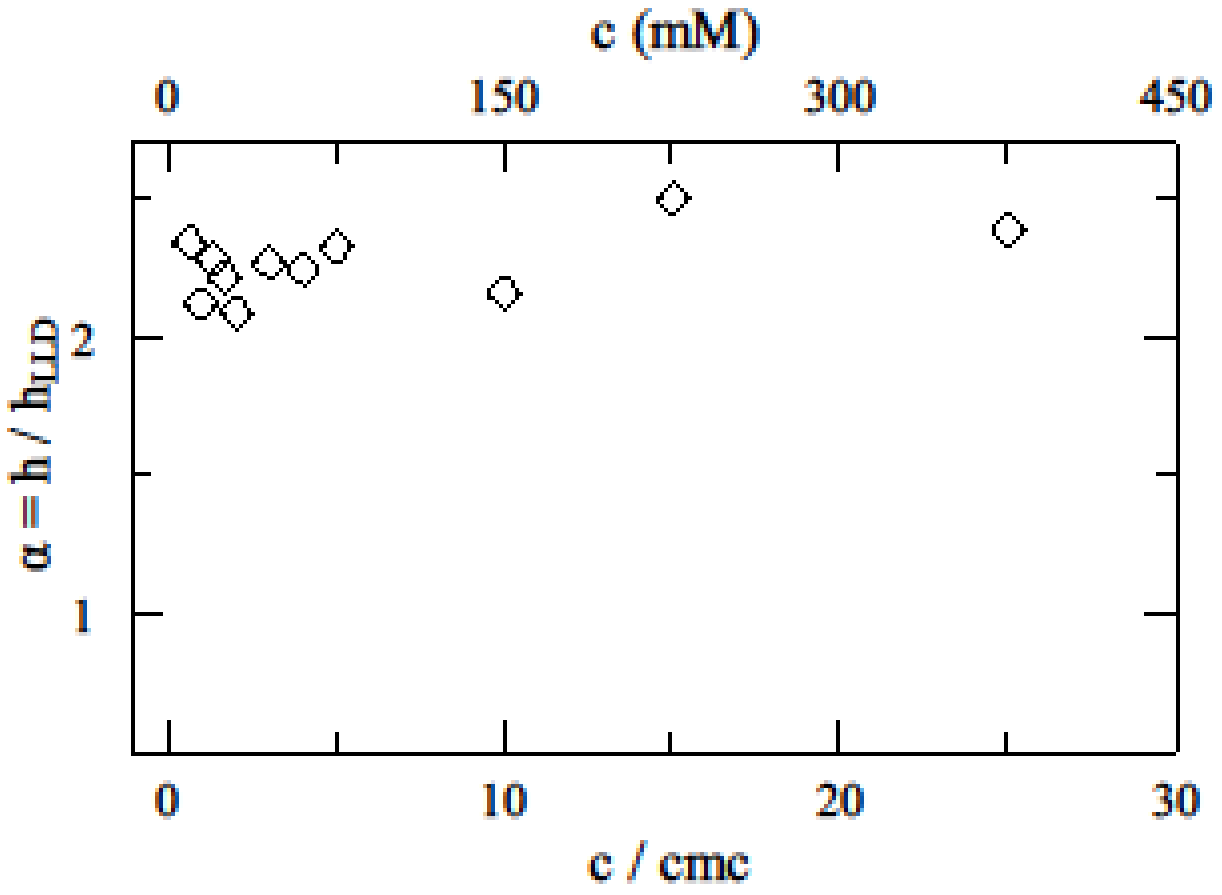

**Figure 8: Thickening factor α for DTAB versus concentration rescaled by the cmc (15 mM).**

### 3.2.2 DeTAB

In order to investigate large surfactant concentrations, we chose to use DeTAB which allows the viscosity to remain small and constant in the entire range of investigated *Ca* values. As shown in Figure 9(a), the beginning of the thickening transition is observed with a 495 mM (i.e., 7.5 cmc) solution. With twice this concentration, 990 mM (i.e., 15 cmc) (Figure 9(b)) the end of the thickening transition is visible. The thickening factor is constant at high *Ca* values and slightly greater than unity (the LLD prediction): $\alpha = 1.06 \pm 0.05$ (Scheid et al. 2010).

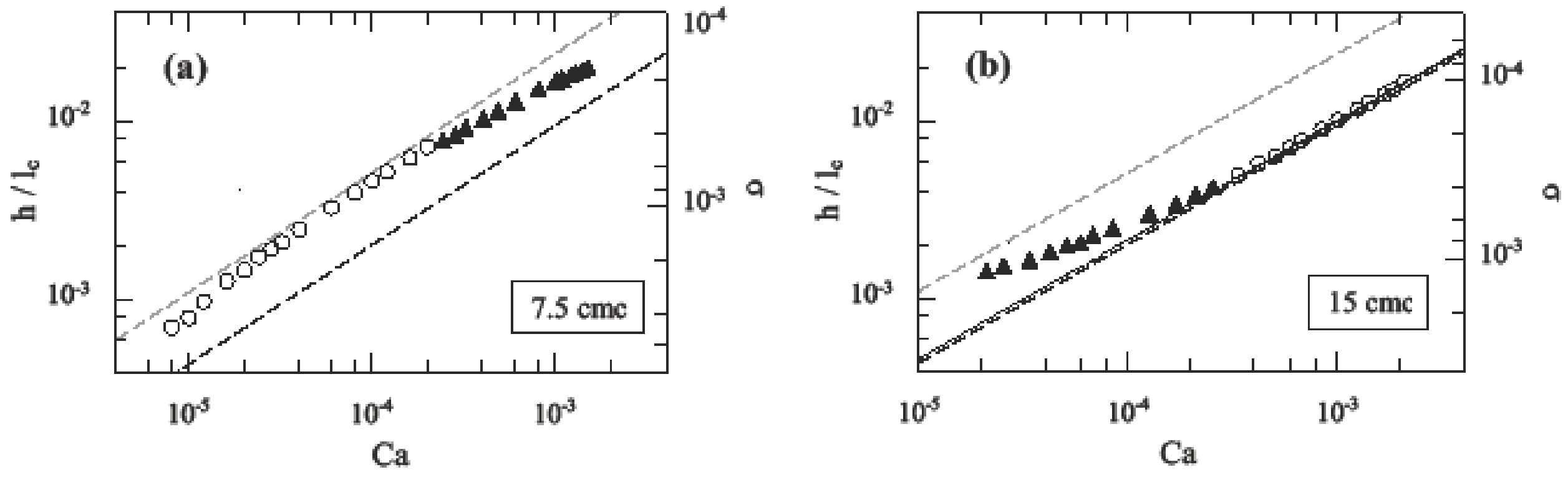

**Figure 9: Results obtained for DeTAB for two concentrations (with a cmc of 66 mM). Same axes and notations as for Figure 5.**

## 3.3 Interpretation

In the following section, we will discuss the observed dynamical transition. As stated in the introduction, a similar transition was already observed in another study in the fiber geometry [4] but no systematic measurements in which surfactants and surfactant concentration are varied in a large range have been performed.

As explained in the introduction, the thickening factor depends on the surface viscoelasticity, hence on the flow characteristics through the capillary number. More specifically, “the dynamic transition of thickening” is linked to the replenishment of the surface by surfactants. Many mechanisms are in competition here: transport of surfactants by diffusion and/or convection towards the interface as well as their adsorption/desorption. Moreover, the film thickness and the surfactants concentration are also very important since a small thickness and/or small concentration can prevent the surfactant remobilization through the confinement effect (see section …)

First of all, let us discuss diffusion and adsorption. One reason why we consistently observe a large thickening may be that surfactants do not have enough time to reach the surface during the experiment. Nevertheless, convection is always efficient enough to replenish the surface (Quéré 1998, Shen 2002): incompressibility of the liquid allows the mass conservation equation $\nabla.\mathbf{u} = 0$, to be approximated as $V/\ell \sim v_t/h$, where $V$, $\ell$, $v_t$ and $h$ are, respectively, the withdrawal velocity, the dynamic meniscus extension, the characteristic transverse velocity in the film and the film thickness. As a result, $h/v_t$ which is the time required to convect surfactants across the entire film is comparable to the time $l/V$ required for the surfactant to go through the entire dynamic meniscus. Diffusion can also play a role for very thin films. The time necessary to diffuse through the film is $h^2/D$ where $D \sim 5\ 10^{-10}$ m²/s is the diffusion coefficient of the surfactants. The convection and diffusion times are of the same order of magnitude when $V \approx \ell D/h^2$ (note that this is equivalent to the calculation of a Péclet number comparing diffusion and convection times), which corresponds to $Ca = ((\eta D)/(\gamma \alpha^3 \ell_c))^{1/2} \approx 3\times 10^{-5}$ (using $\ell \approx l_c \alpha^{1/2} Ca^{1/3}$ and $h \approx \alpha l_c Ca^{2/3}$). So at small capillary numbers, diffusion may play a role while it becomes negligible at large $Ca$. In any case, any surfactant present into the film has enough time to reach the surface during the experiment by either convection or by diffusion. So, the other mechanisms are limiting in our case. The thickness and concentration effects are then discussed in paragraphs 3.3.1 and 3.3.2 whereas the adsorption is discussed in paragraph 3.3.2.

### 3.3.1 Film thickness effects

Let us first focus on apparent surface viscoelasticity variations with film thickness. This point was quantitatively addressed by Lucassen and coworkers (Lucassen 1972, 1981, Lucassen-Reynders 1969). Their model is based on an analogy between a surface perturbed during a very short time, where the surfactant motion is limited by diffusion and a thin film in which surface replenishment is instantaneous. When the surface of a semi infinite solution is perturbed during a short time, surfactants can be exchanged between bulk and surface due to diffusion. The diffusion coefficient of surfactant molecules is assumed to be constant from the bulk up to the surface and the solution dilute enough so that no micelles are present. If the surface of the solution is stretched quickly, diffusion is not fast enough to replenish the interface and the surface viscoelastic moduli are increased. On the contrary, if the expansion is very slow, surfactant concentration gradients are rapidly smoothed out by diffusion of surfactant from the bulk toward the interface, and the surface viscoelastic moduli are decreased.

The refilling efficiency can be quantified by the length scale $(D\tau_e)^{1/2}$ over which surfactants can be remobilized due to diffusion during the expansion time $\tau_e$, where $D$ is the diffusion coefficient of the surfactant. This length scale can be used to relate the thin film surface viscoelastic moduli as a function of thickness to the surface viscoelastic moduli of the bulk solution as a function of expansion time: the surface elasticity of a thin film is the same as that of a solution for an expansion time $\tau_e$, if the film thickness $h$ is equal to $(D\tau_e)^{1/2}$. The surface elastic moduli are small at low frequency (large $\tau_e$) and large at high frequency (small $\tau_e$), with the transition frequency increasing with concentration [11]. Consequently, the film elastic modulus decreases when $h$ increases, as expected from the surface-film analogy (Prins 1967). We then expect a high $\alpha$ at small film thicknesses which is actually observed (Figure 5). Unfortunately, it is impossible to extract a quantitative expression for the thin film elasticity since our experiments are performed at very high concentration at which micelles are present. The relation between $c$ and $\Gamma$ is not known precisely, so we are unable to calculate the surface elastic moduli variations of a solution with respect to time of expansion and concentration and predict the characteristic thickness at which this transition is supposed to occur. However, we carried out systematic measurements of the thickening factor ($\alpha$), obtained before the dynamical transition of thickening (if any), with respect to $C_{12}E_6$ and DTAB concentration (c) (Figure 6 and Figure 8, respectively). In both cases, we observe very reproducible large thickening factors ($\alpha > 2$) for concentration at and above the cmc which shows that the confinement (or thickness) effect is very robust.

### 3.3.2 Transition from large to small thickening

The occurrence of a thickening transition, when increasing solution concentration or film thickness, is a general behavior that was observed with all three surfactants. It had also been observed on withdrawn fibers in the presence of DTAB by Quéré and co-workers (1998). As these authors suggested, the solution must be concentrated enough to allow the refilling of the interface during film formation in the dynamic meniscus. The dynamic meniscus acts as a surfactant reservoir with a thickness that increases with *Ca*. Therefore, for a given concentration, at low film thickness (i.e., low *Ca*), a shortage of surfactants can be expected, resulting in an increase of apparent surface viscoelastic moduli. In turn, beyond a threshold thickness, the reservoir contains enough surfactant to refill the interface, which results in a decrease of the apparent surface viscoelastic moduli. This "confinement effect" is the basis of the calculation of the film elastic modulus of Prins and coworkers discussed in section 3.3.1. The transition observed is then a confinement effect.

As stated in the introduction, Quéré and de Ryck (1998) introduced a dimensionless number $\sigma$ to estimate the capacity of the bulk to act as a surfactant reservoir. This parameter compares the amount of surfactant molecules adsorbed at the interface and present in the dynamic meniscus, and is given by $\sigma = \Gamma/(ch)$. In Figure 5, Figure 7 and Figure 9, the right vertical axes give $\sigma$ calculated with $\Gamma \approx 2$ molecules/nm$^2$, which is a typical value for the surfactants used at c ≥ cmc (Israelachvili 1992). As can be seen in Figure 5, in the case of $C_{12}E_6$ the thickening transition occurs when $\sigma$ is of order unity, which is when the amount of surfactant molecules in the dynamic meniscus is of the same order of magnitude as the amount of adsorbed molecules. In the case of DTAB (see Figure 7), the thickening transition occurs, more surprisingly, around $10^{-3}$, suggesting that here the "confinement effect" (for small film thickness) is not the only effect involved as discussed in the next section.

### 3.3.3 Adsorption barrier effects

In the case of ionic surfactants, the thickening transition occurs at higher *Ca* and *h* values and for higher bulk concentrations ($\sigma << 1$). Let us note that the dynamic transition of thickening observed by Ou Ramdane and Quéré (1998) for DTAB also occurred well below $\sigma = 1$. This response could be due to adsorption electrostatic barriers associated with the charged surfactant monolayers present at the surface. Such a barrier indeed leads to an increase of the time necessary for the surfactants to reach the surface by an exponential factor $\exp(W/k_BT)$, where $W$ is the adsorption energy barrier, $k_B$, the Boltzmann constant and $T$ the absolute temperature. For DTAB close to the cmc, W ~ 15 $k_BT$ (Ritacco 2011). Addition of salt lowers the energy barrier (electrostatic screening), which disappears above salt concentrations of about 100mM. Addition of large amounts of ionic surfactant produces a similar self screening effect, which is expected to lead to the disappearance of the barrier as well. The effective compression elastic modulus of dilute ionic surfactant solutions is much larger than predicted by the Lucassen-van den Tempel model at high surfactant concentrations, but decreases and tends towards the values predicted by the model when enough salt is added (Bonfillon 1994). This observation could account for the fact that the thickening transition is observed for much larger surfactant bulk concentrations in the case of ionic surfactants than with nonionic surfactants.

We then propose that, below 100mM, even though the surfactants are available at high enough concentration and have time to reach the surface, the electrostatic barrier prevents them from adsorbing. So, as soon as $W$ is large enough to prevent adsorption, it is reasonable to assume that the surfactants cannot adsorb at the interface, leading to large effective elasticity and then a large thickening factor. Now, this effect of electrostatic barrier decreases with increasing surfactant concentration (like for the salt in the example above).

### 3.3.4 Concentration effects

In this last section, we look at the concentration effects on the dynamical transition of thickening, focusing on the experimental results with $C_{12}E_6$ for which no absorption barrier effects are expected. Provided the thickening transition is due to the confinement effects gauged by the parameter $\sigma$ (see section 3.3.1 and 3.3.2), we report in Figure 10 the value of this parameter at the beginning of the transition, denoted $\sigma_{tr}=\Gamma/(ch_{tr})$, versus the scaled concentration $c$/cmc, with $h_{tr}$ the thickness at the transition (see Figure 5 (b)). The logarithmic fit of the data represented by the solid line in Figure 10 gives

$$\sigma_{tr} = \frac{\sigma_0}{c/cmc}, \qquad (4)$$

where $\sigma_0 \simeq 12.4$ is the fitting constant. Assuming $\sigma_0$ to be independent of the concentration down to the cmc, we can write $\sigma_0=\Gamma/(cmc \cdot h_{tr})$. Taking a typical value of $\Gamma$ = 1 molecule/50Å$^2$ (*i.e.*, $\Gamma \simeq 3\times10^{-6}$mol/m$^2$), with cmc = 0.07 mM, we find $h_{tr} \simeq 3.5$ µm, which matches the observations in Figure.5. Now, using the surface-film analogy, we can extract a typical frequency equivalent to this value of the film thickness at the transition: $f_{tr} = D/h_{tr}^2 \simeq 40$ Hz, where $D = 5\times10^{-10}$ m$^2$/s.

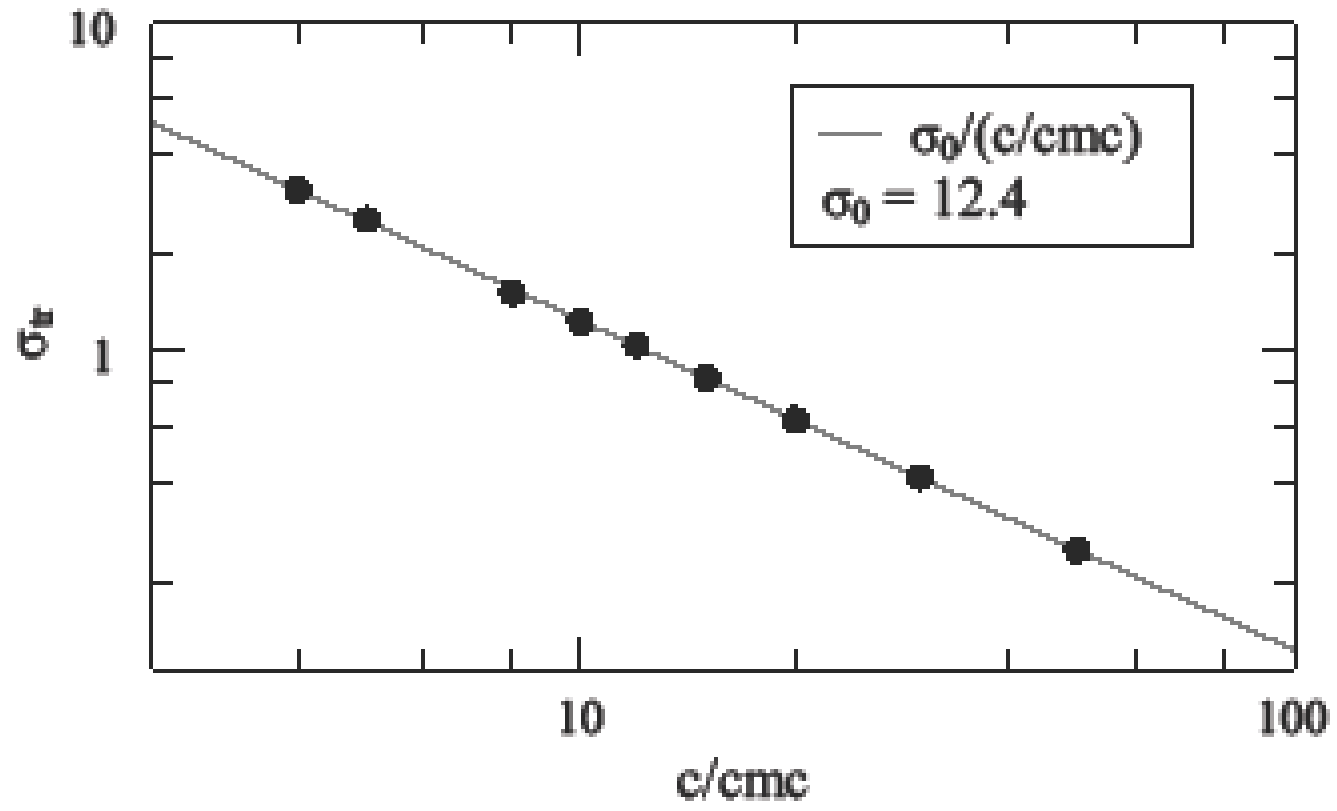

**Figure 10: Values of the parameter at the beginning of the transition $\sigma = \sigma_{tr}$ (see Figure 5 (b)) as a function of concentration of $C_{12}E_6$ scaled by cmc, and fit (solid line) with equation (4).**

Our experiments then show that, at a frequency higher than 40Hz (i.e. at a thickness smaller than 3.5 µm), the surface elasticity saturates at a high value. Some experiments have been done by Stubenrauch et al. (2009) to measure directly the surface elasticity of $C_{12}E_6$ around the cmc at small frequencies (bellow 1Hz). The authors extrapolate their results at high frequency and observe a saturation of the surface elasticity around few tens of Hertz. We think that what we observe in our experiment corresponds to this saturation of surface elasticity at high frequency.

Unfortunately, to our knowledge, no measurements at concentrations higher than the cmc and at high frequency are available in literature. These measurements, using capillary waves to reach high frequencies, are beyond the scope of this paper.

The transition from large to small thickening is determined by the bulk concentration of surfactant.. In this "classical" transport picture, the micelle disassembly provides a "source" of monomer and the transport of the surfactants into or out of the micelle is not taken into account. Due to their small size, surfactants diffuse to the surface much faster than the micelle and, in the case of low micelle concentration a depletion zone can appear. The concentration of surfactant at the interface then depends on micelle break-down kinetics. Maldarelli and coworkers (Bhole 2010, Song 2002) have studied this effect both experimentally and theoretically. At large surfactant concentrations (as in our experiment), this effect can be neglected as micelle breakdown is extremely fast.

# 4. Conclusion

This work provides an extensive experimental study of solid coating in the plate geometry while varying the type of surfactant and concentration. Our experiments show two main features:

- First, we confirmed repeatedly the thickening of the withdrawn liquid film with respect to the LLD prediction for every concentration and in the entire range of *Ca* for which gravity is negligible.
- Second, we provided evidence of two regimes separated by a dynamical thickening transition. At small capillary number, the film thickness is very small and a confinement effect can be observed. Then, at a thickness predicted by the dimensionless number $\sigma$ for nonionic surfactants, the confinement effect disappears and $\alpha$ slowly evolves towards the LLD prediction. At very high concentration and thickness, this transition ends and a small remaining thickening is observed, due to intrinsic surface viscosity, as explained in our previous paper ([13].

The thickening effects and transitions are driven by interfacial rheological properties coupled to surface refilling mechanisms. Moreover, thickening transitions seem to be a general feature in thin films made of surfactant solutions. Elasticity of the interface is the main source of thickening in the case of interfacial surfactant concentration gradients, as it is the case for either low concentrations or films with low thickness, whereas, for intermediate concentrations or thicker films, interfacial viscosity also plays a role.

For non ionic surfactants, the thickening transition occurs for $\sigma$ of order unity. This feature is not true anymore for the ionic surfactants and we suggest that this behavior is due to an electrostatic barrier that prevents the adsorption of surfactants at small concentration and thus shifts the thickening transition towards higher bulk concentrations.

**Acknowledgements** We are grateful to Isabelle Cantat and Ernst van Nierop for numerous helpful discussions. We also thank François Boulogne for valuable help with experimental set-up. BS acknowledges the Brussels region for financial support through the program “Brains Back to Brussels” as well as the FRS-FNRS.

.